\documentclass[11pt]{article}
\usepackage{fullpage,epsf,amssymb,amsthm,amsfonts,amsmath,latexsym, graphicx,array,extarrows,mathtools,mathstyle,mathrsfs,slashed,subcaption,amsbsy,csquotes}
\usepackage[normalem]{ulem}
\usepackage{authblk}
\usepackage{cite}
\usepackage{color}

\newcommand{\ud}{\mathrm{d}}
\newcommand{\uvec}[1]{\boldsymbol{#1}}

\title{\bf{Analytic constraints on the energy-momentum tensor in conformal field theories}}

\author[1]{C\'{e}dric Lorc\'{e}\thanks{cedric.lorce@polytechnique.edu}}
\author[1]{Peter Lowdon\thanks{peter.lowdon@polytechnique.edu}}

\affil[1]{\footnotesize{\textit{CPHT, CNRS, Ecole Polytechnique, Institut Polytechnique de Paris, Route de Saclay, 91128 Palaiseau, France}}}

\date{}
\begin{document}

{\let\newpage\relax\maketitle}
\setcounter{page}{1}
\pagestyle{plain}

\abstract
\noindent
In this work we investigate the matrix elements of the energy-momentum tensor for massless on-shell states in four-dimensional unitary, local, and Poincar\'e covariant quantum field theories. We demonstrate that these matrix elements can be parametrised in terms of covariant multipoles of the Lorentz generators, and that this gives rise to a form factor decomposition in which the helicity dependence of the states is factorised. Using this decomposition we go on to explore some of the consequences for conformal field theories, deriving the explicit analytic conditions imposed by conformal symmetry, and using examples to illustrate that they uniquely fix the form of the matrix elements. We also provide new insights into the constraints imposed by the existence of massless particles, showing in particular that massless free theories are necessarily conformal.

\newpage

\section{Introduction}
\label{intro}

As with any quantum field theories (QFTs), the correlation functions in conformal field theories (CFTs) completely encode the dynamics of these theories. A characteristic property of CFTs is that the conformal symmetry significantly constrains the form of the correlation functions, reducing the classification of these objects to the determination of a series of constant parameters. Although the analysis of CFTs has historically focussed on the Euclidean version of these theories, in part because of their relative analytic simplicity, in recent years there has been a surge in interest in Minkowskian CFTs, particularly in the context of the analytic bootstrap program~\cite{Poland:2018epd}. Due to the larger number of space-time symmetries, the energy-momentum tensor (EMT) plays a central role in many of the analytic constraints imposed on CFTs. An important example are the three-point functions involving the EMT, which have been shown in Euclidean space~\cite{Osborn:1993cr,Erdmenger:1996yc,Bzowski:2013sza,Bzowski:2018fql}, and more recently for certain cases in Minkowski space~\cite{Bautista:2019qxj}, to be fully constrained by the conserved conformal currents, and their corresponding Ward identities. \\

\noindent
The focus of this work will be the Minkowski matrix elements of the EMT for massless on-shell states. In particular, we will study these matrix elements in QFTs that are unitary, local, and Poincar\'e covariant. By local we mean that all of the fields $\Phi(x)$ in the theory, including the EMT, either commute or anti-commute with each other for space-like separations. Poincar\'e covariance implies that the components of these fields $\Phi_{k}(x)$ transform as:
\begin{align}
U(a, \alpha)\Phi_{k}(x)U^{-1}\!(a, \alpha) = \sum_{l}\mathcal{D}_{kl}^{(\Phi)}(\alpha^{-1}) \Phi_{l}(\Lambda(\alpha)x + a),
\label{Poincare_tran}
\end{align} 
under (proper orthochronous) Poincar\'e transformations $(a,\alpha)$, where $\mathcal{D}^{(\Phi)}$ is the corresponding Wigner matrix that defines the representation of the field, and $\Lambda(\alpha)$ is the four-vector representation of $\alpha$~\cite{Haag:1992hx}. Although the overall structural properties of EMT matrix elements have been understood for many years, at least in the case of massive states~\cite{Pagels:1966zza,Boulware:1974sr}, the explicit spin dependence of these objects has only been studied relatively recently. By using the conservation of the EMT, together with its various symmetry properties, one can decompose these matrix elements into a series of form factors with independent covariant coefficients\footnote{By \textit{covariant} we mean that the components of the coefficients transform in the same manner as the fields [Eq.~\eqref{Poincare_tran}] under Lorentz transformations.}. We refer to these throughout as the gravitational form factors (GFFs). As the spin of the states increases, these objects become increasingly more complicated due to the larger number of potential covariant structures. This explains in part why many studies\footnote{See~\cite{Donoghue:2001qc,Maybee:2019jus} and~\cite{Ji:1996ek,Polyakov:2002yz,Lorce:2017wkb,Polyakov:2018zvc,Lorce:2018egm,Cosyn:2019aio} for some recent examples of gravitational and hadronic studies.} have focussed on calculations for states with lower values of spin, generally $0$, $\tfrac{1}{2}$, or $1$. Once the potential covariant coefficients are known, constraining the corresponding GFFs is therefore essential for understanding the analytic structure of the matrix elements. Since the states appearing in the on-shell matrix elements are definite momentum eigenstates, and hence not normalisable, this has led to incorrect physical conclusions in the literature, as detailed in~\cite{Bakker:2004ib}. This non-normalisability stems from the fact that the matrix elements are distributions, not functions. It was first shown for the massive spin-$\tfrac{1}{2}$ case that by taking this property into account in the derivation of the GFF constraints, these apparent ambiguities no longer arise~\cite{Lowdon:2017idv}. This approach was later generalised to massive states with arbitrary values of spin~\cite{Cotogno:2019xcl}, as well general spin-state definitions, including massless states~\cite{Lorce:2019sbq}. \\

\noindent
Most studies of the CFT three-point functions involving the EMT have focussed on cases where the other fields have lower values of absolute helicity $|h|$, since increasing $|h|$ quickly leads to complicated expressions. Because these three-point functions are directly related to the massless on-shell EMT matrix elements, via a projection of the Lorentz components of the fields, it turns out that this increase in complexity is simply a different realisation of the fact that the number of independent covariant structures in the form factor expansion increases with $|h|$. Now whilst it is clear that the total number of these independent structures must be finite for different values of $|h|$, establishing what these numbers are is non-trivial. In a recent work~\cite{Cotogno:2019vjb}, this problem was solved for the EMT matrix elements of massive states with arbitrary spin $s$. The essential step in this analysis was the realisation that \textit{all} covariant structures that can appear in the matrix elements can be generated by contracting the covariant multipoles\footnote{For example, $\mathcal{M}_{0}=1$, $\mathcal{M}_{1}^{\mu\nu} \!= S^{\mu\nu}$, and $\mathcal{M}_{2}^{\mu\nu,\rho\sigma} \!= \tfrac{1}{2}\!\left\{S^{\mu\nu}, S^{\rho\sigma} \right\} -\tfrac{1}{12}\!\left(g^{\mu\rho}g^{\nu\sigma}- g^{\mu\sigma}g^{\nu\rho}\right)S\cdot S +\tfrac{1}{4!}\epsilon^{\mu\nu\rho\sigma}\epsilon_{\alpha\beta\gamma\delta}S^{\alpha\beta}S^{\gamma\delta}$ define the monopole, dipole, and quadrapole, respectively. See~\cite{Cotogno:2019vjb} for an in-depth discussion of these objects.} $\{\mathcal{M}_{n}\}$ of the Lorentz generators $S^{\mu\nu}$ with the external momenta $p'$ and $p$, and the metric. The advantage of using this multipole basis is that these objects explicitly truncate for each value of $s$. In particular, given states of spin $s$, one has the constraint:
\begin{align}
\mathcal{M}_{N}=0, \quad  \text{for} \ \ N>2s,
\label{multi_bound}
\end{align}
where $\{\mathcal{M}_{N}\}$ are constructed from Lorentz generators that transform under the same representation as the fields creating the states. Not only does this constraint prove that only a finite number of independent covariant structures enter into the EMT matrix elements, it also provides a basis from which these structures can be systematically characterised~\cite{Cotogno:2019vjb}. Since the covariant multipoles are fundamental objects that exist independently of the specific properties of the states, this representation can also equally be applied to the matrix elements of massless states~\cite{Lorce:2019sbq}. As will be outlined in Sec.~\ref{grav_FF}, this has the important implication that the helicity dependence of these matrix elements factorises, or equivalently, that the dependence on the Lorentz representation of EMT three-point functions can be written in a manifest way. \\ 

\noindent
Before concluding this section we will first discuss a result which plays a central role throughout this work, the \textit{Weinberg-Witten Theorem}~\cite{Weinberg:1980kq}. This theorem puts constraints on the potential structure of matrix elements involving massless on-shell states, in particular implying that:
\begin{align}
\langle p',h'|T^{\mu\nu}(0)|p ,h \rangle =0, \quad \quad \text{for} \ \ |h'+h| \neq 0, 1, 2 \ \ \text{and} \ \ (p'-p)^{2}<0.
\label{WW}
\end{align} 
In~\cite{Weinberg:1980kq} it is further stated that this constraint can be extended by continuity to the point $p'=p$. It turns out though that potential discontinuities can in fact occur when $p'=p$ due to the distributional nature of the matrix elements~\cite{Sudarshan:1981cj}. However, by making the additional assumption that the corresponding QFT is a \textit{local} theory, this argument can be made consistent~\cite{Lopuszanski:1983zx}. Since the EMT operator does not modify the value of $|h'|$ or $|h|$, the constraints from Eq.~\eqref{WW} subsequently lead to the important conclusion:
\begin{displayquote}
\textit{Massless particles of helicity $h$, where $|h|>1$, cannot possess charges induced by a local and Poincar\'e covariant energy-momentum tensor.}
\end{displayquote}
As emphasised in~\cite{Lopuszanski:1983zx}, this does not mean that consistent Poincar\'e or conformal generators do not exist for massless states with higher helicity, only that these generators cannot be written in terms of integrals of a local and Poincar\'e covariant EMT. In other words, by allowing massless particles with $|h|>1$ this necessarily requires that the corresponding EMT is either non-local, non-covariant, or both. This imposes significant constraints on the structure of massless EMT matrix elements, as will be discussed in Sec.~\ref{grav_FF}.  \\

\noindent
The remainder of this paper is structured as follows. In Sec.~\ref{grav_FF} we combine the covariant multipole approach with constraints from the Weinberg-Witten Theorem to derive a general form factor decomposition for the EMT matrix elements of massless on-shell states, and in Sec.~\ref{conf_FC} we go on to outline some important properties of conformal fields and currents which we will need for the calculations in subsequent sections. Using the results from Sec.~\ref{grav_FF} and~\ref{conf_FC}, in Sec.~\ref{matrix_constr} we derive the GFF constraints imposed by conformal symmetry and the trace properties of the EMT, and apply these findings to specific CFT examples. In Sec.~\ref{massless_part} we combine the results derived throughout the paper to establish a general connection for free theories between the existence of massless particles and the presence of conformal symmetry, and finally in Sec.~\ref{concl} we summarise our key findings.

\section{Gravitational form factors for massless states}
\label{grav_FF}

For the purposes of this paper we are interested in the EMT matrix elements for on-shell momentum eigenstates. One can covariantly impose this on-shell restriction by defining states with mass $M$:
\begin{align}
|p, \sigma ;M \rangle = \delta_{M}^{(+)}(p)|p, \sigma\rangle = 2\pi \,\theta(p^{0})\,\delta(p^{2}-M^{2}) |p, \sigma\rangle,
\label{M_state}
\end{align}
where $|p, \sigma\rangle$ is the standard non-covariant momentum eigenstate\footnote{The difference between these states is that the on-shell factor is included in the definition of $|p, \sigma ;M \rangle$, as opposed to the momentum integration measure. In constructing physical states one therefore integrates $|p, \sigma ;M \rangle$ over $\frac{d^{4}p}{(2\pi)^{4}}$, whereas for $|p, \sigma\rangle$ the measure itself is on shell: $\frac{d^{4}p}{(2\pi)^{4}}\delta_{M}^{(+)}(p) = \frac{d^{3}\uvec{p}}{(2\pi)^{3}2E_{p}}dp^{0}\delta(p^{0}- E_{p})$, with $E_{p} = \sqrt{\uvec{p}^{2} + M^{2}}$.} with internal quantum numbers $\sigma$. The advantage of using the states in Eq.~\eqref{M_state} is that they transform covariantly under Poincar\'e transformations, which significantly simplifies on-shell matrix element calculations. In what follows, we will restrict ourselves to massless on-shell states. For simplicity we drop the label $M=0$ and denote these states by $|p,h\rangle$, where $h$ is the helicity. Since we focus only on unitary QFTs throughout this work, it follows from Eq.~\eqref{M_state} and the standard inner product for general eigenstates\footnote{In particular: $\langle p',\sigma'|p ,\sigma \rangle = 2p^{0}\,(2\pi)^{3}\delta^{3}(\uvec{p}'-\uvec{p})\,\delta_{\sigma'\sigma}$.} that:
\begin{align}
\langle p',h'|p ,h \rangle = (2\pi)^{4}\delta^{4}(p'-p)\,\delta_{0}^{(+)}\!(p)\,\delta_{h'h}.
\label{norm}
\end{align}   
In a previous analysis, which explored the GFF constraints imposed by Poincar\'e symmetry~\cite{Lorce:2019sbq}, it was established under the assumptions that the EMT is symmetric, hermitian, and both parity $\mathsf{P}$ and time-reversal $\mathsf{T}$ invariant, that the EMT matrix elements for on-shell massless states in a unitary, local, Poincar\'e covariant QFT have the following general decomposition:
\begin{align}
\langle p',h'|T^{\mu\nu}(0)|p ,h \rangle &= \overline{\eta}_{h'}(p')\Big[\bar{p}^{\{\mu}\bar{p}^{\nu\}}  A(q^{2}) + i \bar{p}^{\{\mu}S^{\nu\}\rho}q_{\rho} \, G(q^{2})    + \cdots  \Big]\eta_{h}(p)  \, \delta_{0}^{(+)}\!(p')\,\delta_{0}^{(+)}\!(p),
\label{T_decomp_0}
\end{align}
where $\cdots$ represent contributions with an explicitly higher-order dependence on the four-momentum transfer $q=p'-p$, index symmetrisation is defined: $a^{\{\mu }b^{\nu\}}= a^{\mu}b^{\nu}+a^{\nu}b^{\mu}$, and $\bar{p}= \tfrac{1}{2}(p'+p)$. We refer to $\eta_{h}(p)$ as the generalised polarisation tensors (GPTs), which correspond to the Lorentz representation index-carrying coefficients appearing in the decomposition of primary free fields with helicity $h$, with normalisation chosen such that: $\overline{\eta}_{h'}(p)\eta_{h}(p) = \delta_{h'h}$. For example, since we assume $\mathsf{P}$ and $\mathsf{T}$ invariance, the GPT in the $|h|=\tfrac{1}{2}$ case is proportional to the Dirac spinor $u_{h}(p)$. As discussed in Sec.~\ref{intro}, one can see in Eq.~\eqref{T_decomp_0} that this expression is constructed by contracting covariant multipoles, in this case the monopole and dipole, with all possible combinations of momenta and the metric. Due to the low number of covariant indices, it turns out that up to linear order in $q$ there exists only one such combination for each multipole which is consistent with the various symmetries of the EMT\footnote{This characteristic is also true for massive states~\cite{Cotogno:2019xcl}.}. As detailed in~\cite{Lorce:2019sbq}, the constraints arising from Poincar\'e symmetry\footnote{Due to Poincar\'e symmetry one finds that: $A(q^{2})  \delta^{4}(q)  = \delta^{4}(q)$, and: $G(q^{2})  \delta^{4}(q)  = \delta^{4}(q)$~\cite{Lorce:2019sbq}.} only affect the GFFs with coefficients that depend at most linearly on $q$, which explains why only these leading terms are considered in Eq.~\eqref{T_decomp_0}. However, in general there are other possible GFFs, with coefficients that potentially involve contractions with higher multipoles. For massive states, these GFFs were fully classified in~\cite{Cotogno:2019vjb} for arbitrary spin. In the remainder of this section we will discuss the massless case. \\

\noindent
In general, given a massless field with Lorentz representation $(m,n)$, this field creates states with helicity $h=n-m$. For example, a left-handed Weyl spinor with representation $(\tfrac{1}{2},0)$ gives rise to $h=-\tfrac{1}{2}$ states. Massless representations of the Lorentz group have significant additional constraints imposed upon them, including the fact that \textit{all} irreducible representations can be built from representations $(m,n)$ where either $m=0$, $n=0$, or both~\cite{Lopuszanski:1991}. In particular, it follows that massless fields which transform covariantly under the vector representation $(\tfrac{1}{2},\tfrac{1}{2})$ cannot be irreducible. This can be explicitly seen by the fact that any such field\footnote{This constraint gives rise to the well-known result that the components of the massless photon field $A_{\mu}$ cannot transform covariantly as a vector without violating the positivity of the Hilbert space (unitarity)~\cite{Nakanishi:1990qm}. We will discuss this characteristic further in Sec.~\ref{higher_h}.} $V_{\mu}$ can always be written as the gradient of a scalar field: $V_{\mu} = \partial_{\mu}\phi$, since $V_{\mu}$ only defines states with $h=0$. More generally, \textit{any} massless field with Lorentz representation $(m,n)$, where both $m$ and $n$ are non-vanishing, can be written in terms of derivatives of irreducible fields~\cite{Lopuszanski:1991}. Since massless QFTs are constructed from irreducible fields, or their direct sums, the corresponding GPTs of these fields, including those in Eq.~\eqref{T_decomp_0}, must also transform under these representations. As we will see, this puts significant constraints on the type of GFFs that can appear in the EMT matrix elements. \\

\noindent
In Sec.~\ref{intro} we introduced the Weinberg-Witten Theorem and outlined its implications, namely that if massless particles with $|h|>1$ exist, it follows that the corresponding EMT must either be non-local, non-covariant, or both. For the EMT matrix elements with $|h|>1$, this implies that these objects must necessarily contain a form factor with either a non-local or non-covariant coefficient. However, since we restrict ourselves  in this work to unitary CFTs which are local and Poincar\'e covariant, these matrix elements must in fact vanish~\cite{Lopuszanski:1983zx}. In other words, Eq.~\eqref{T_decomp_0} is only non-trivial when the states have helicity $|h|\leq 1$. Another constraint on the type of GFFs appearing in Eq.~\eqref{T_decomp_0} comes from the covariant multipole bound in Eq.~\eqref{multi_bound}. If the GPTs are in the Lorentz representation $(m,n)$, it equally follows in the massless case that the multipoles $\{\mathcal{M}_{N}\}$ must vanish for $N>2(m+n)$. Since by definition $\mathcal{M}_{N}$ contains $N$ products of the Lorentz generators $S^{\mu\nu}$, which transform under the same (irreducible) representation as the GPTs: $(m,0)$, $(0,n)$, or their direct sums, the number of powers of these generators is therefore bounded above by the \textit{helicity} of the states. In particular:
\begin{displayquote}
\textit{The number of powers of $S^{\mu\nu}$ appearing in the EMT matrix elements for massless states of helicity $h$ is at most $2|h|$.}
\end{displayquote}
Combining this with the Weinberg-Witten Theorem constraint $|h|\leq 1$, one is immediately led to the conclusion that only GFFs which have coefficients with \textit{two or fewer} powers of $S^{\mu\nu}$ are permitted. Now we are in a position to write down the full generalisation of Eq.~\eqref{T_decomp_0}. If we continue to demand that the GFFs are dimensionless, as in Eq.~\eqref{T_decomp_0}, and also similarly assume that the EMT is symmetric, hermitian, and both $\mathsf{P}$ and $\mathsf{T}$ invariant, the non-trivial ($|h|\leq 1$) matrix elements of the EMT for massless on-shell states have the general form\footnote{In particular, for $|h|=\tfrac{1}{2}$ the coefficient of $T(q^{2})$ is no longer independent, and so only $A(q^{2})$, $G(q^{2})$ and $C(q^{2})$ can potentially exist, and for $h=0$, since $(S^{\mu\nu})_{(0,0)}=0$, only $A(q^{2})$ and $C(q^{2})$ remain. }:
\begin{align}
\langle p',h'|T^{\mu\nu}(0)|p ,h \rangle &= \overline{\eta}_{h'}(p')\Big[\bar{p}^{\{\mu}\bar{p}^{\nu\}}  A(q^{2}) + i \bar{p}^{\{\mu}S^{\nu\}\rho}q_{\rho} \, G(q^{2})  \nonumber \\
& \quad  + 2(q^{\mu}q^{\nu} - q^{2}g^{\mu\nu}) \, C(q^{2}) + S^{\{\mu\alpha}S^{\nu\}\beta}q_{\alpha}q_{\beta} \, T(q^{2}) \Big]\eta_{h}(p)  \, \delta_{0}^{(+)}\!(p')\,\delta_{0}^{(+)}\!(p).
\label{T_decomp}
\end{align}
From Eq.~\eqref{T_decomp} one can see that by adopting a parametrisation that uses the covariant multipoles as its basis, this leads to a form factor decomposition of the EMT matrix elements in which the helicity dependence is factorised: only knowledge of the generator $S^{\mu\nu}$ in the Lorentz representation of the GPTs is required to calculate the matrix elements for states of different helicities. It is also interesting to note that by virtue of the fact that $q^{2}$ is the only massive Lorentz invariant that can appear in massless theories, there cannot exist form factor coefficients other than those in Eq.~\eqref{T_decomp} which are compatible with locality, whilst also ensuring the form factors are dimensionless. In the remainder of this paper we will derive some of consequences of Eq.~\eqref{T_decomp}.

\section{Conformal fields and currents}
\label{conf_FC}

Besides Poincar\'e symmetry, CFTs are also invariant under infinitesimal dilations and special conformal transformations (SCTs). In the case of dilations, the dilation operator $D$ acts on conformal fields in the following manner:
\begin{align}
i\left[D,\Phi(x)\right] = \left(x^{\mu}\partial_{\mu} + \Delta \right)\Phi(x),
\label{D_transform}
\end{align}
where $\Delta$ is the conformal dimension of the field. For SCTs, the charge $K^{\mu}$ instead has the action:
\begin{align}
i\left[K^{\mu},\Phi(x)\right] = \left( 2x^{\mu}x_{\nu}\partial^{\nu} - x^{2}\partial^{\mu} + 2x^{\mu} \Delta - 2i x_{\nu}S^{\mu\nu}\right) \Phi(x),
\label{SC_transform}
\end{align}
where the Lorentz representation indices of both the field and the Lorentz generator $S^{\mu\nu}$ have been suppressed for simplicity. It turns out that these transformations impose significant constraints on the properties of the fields. In particular, combining Eq.~\eqref{SC_transform} with the masslessness of the field implies the following important relation~\cite{Lopuszanski:1991}:
\begin{align}
(\Delta -1)\partial^{\mu}\Phi(x) = i S^{\mu\nu}\partial_{\nu}\Phi(x).
\label{conformal_field}
\end{align}   
For scalar fields $(S^{\mu\nu})_{(0,0)}=0$, and hence $\Delta=1$; for spinor fields in the $(\tfrac{1}{2},0)$ and $(0,\tfrac{1}{2})$ representations one recovers the Weyl equations when $\Delta=\tfrac{3}{2}$; and for the anti-symmetric tensor field the substitution of $(S^{\mu\nu})_{(1,0)\oplus (0,1)}$ results in the Maxwell equations for $\Delta=2$. In Eq.~\eqref{T_decomp} the EMT matrix element is expressed in terms of the action of Lorentz generators on massless GPTs. By inserting the plane-wave expansion for a massless field into Eq.~\eqref{conformal_field}, one obtains a constraint on this action: 
\begin{align}
(\Delta -1)p^{\mu}\eta_{h}(p) = ip_{\nu} S^{\mu\nu}\eta_{h}(p),
\label{conformal_GPT}
\end{align}  
where the Lorentz representation indices have again been suppressed for simplicity. \\

\noindent
Before discussing the specific structure of the currents associated with dilations and SCTs, it is important to first outline the additional constraints imposed on the EMT itself. In any QFT it is well known that the EMT is not unique since one can always add a \textit{superpotential term}, a term that is separately conserved, but when integrated reduces to a spatial divergence, and hence does not contribute to charges. A prominent example is the pure-spin term that symmetrises the canonical EMT. Superpotential terms also play a particularly important role in CFTs\footnote{See~\cite{Hudson:2017xug} for a further discussion on the relevance of these terms in the context of CFTs.}, since the condition for a theory to be conformal is related to whether or not there exists such a term, which when added to the EMT, renders it traceless. In particular, given a four-dimensional QFT with a conserved and symmetric EMT, $T^{\mu\nu}_{\!(\text{S})}$, a necessary and sufficient condition for this theory to be conformal is that there exists a \textit{local} operator $L^{\mu\nu}(x)$ such that~\cite{Polchinski:1987dy}:
\begin{align}
T^{\mu}_{\!(\text{S})  \mu}(x) = \partial_{\alpha}\partial_{\beta}L^{\alpha\beta}(x).
\label{trace_S}
\end{align}
If this condition holds, it follows that there exists conserved dilation $J^{\mu}_{\!(\text{S})  D}$ and SCT currents $J^{\mu}_{\!(\text{S})  K^{\nu}}$ with the form\footnote{Further background from the early literature on this subject can be found for example in~\cite{Wess:1960,Callan:1970ze,Coleman:1970je}.}:
\begin{align}
&J^{\mu}_{\!(\text{S})  D}= x_{\nu}T_{\!(\text{S})}^{\mu\nu} - \partial_{\nu}L^{\nu\mu}, \label{D_form} \\
&J^{\mu}_{\!(\text{S})  K^{\nu}}= (2x^{\nu}x_{\alpha} - g^{\nu}_{\ \alpha}x^{2})T_{\!(\text{S})}^{\mu\alpha} - 2x^{\nu}\partial_{\alpha}L^{\alpha\mu} + 2 L^{\mu\nu}. \label{K_form}
\end{align}
Under the further assumption that the CFT is unitary, one has that: $L^{\mu\nu}(x) = g^{\mu\nu}L(x)$, and hence the EMT trace condition becomes\footnote{See~\cite{Dymarsky:2013pqa} and references within.}
\begin{align}
T^{\mu}_{\!(\text{S})  \mu}(x) = \partial^{2}L(x).
\label{conformal_constr}
\end{align}
In the general case that Eq.~\eqref{trace_S} is satisfied, this implies that there exists superpotential terms, which when added to Eqs.~\eqref{D_form} and~\eqref{K_form} transform these expressions into the form:
\begin{align}
&J^{\mu}_{\!(\text{ST})  D}= x_{\nu}T_{\!(\text{ST})}^{\mu\nu}, \label{D_current} \\
&J^{\mu}_{\!(\text{ST})  K^{\nu}}= (2x^{\nu}x_{\alpha} - g^{\nu}_{\ \alpha}x^{2})T_{\!(\text{ST})}^{\mu\alpha},  \label{K_current}
\end{align}
where $T_{\!(\text{ST})}^{\mu\nu}$ is both symmetric and traceless. For the remainder of this work we will refer to $T_{\!(\text{ST})}^{\mu\nu}$ as the \textit{modified} EMT\footnote{Particularly in the CFT literature, the symmetric-traceless EMT is often referred to as the \textit{improved} EMT~\cite{Callan:1970ze}.}.

\section{Conformal EMT matrix element constraints}
\label{matrix_constr}

In~\cite{Cotogno:2019vjb, Lorce:2019sbq} Poincar\'e covariance was used to derive constraints on the GFFs for both massive and massless states. In this section we derive the corresponding GFF constraints imposed by conformal covariance, as well as from the trace properties of the EMT itself.

\subsection{Dilational covariance}
\label{dilational}

By definition, momentum space fields have the following action on the vacuum state: $\langle 0|\widetilde{\Phi}(p') = \sum_{h''} \eta_{h''}(p')\langle p',h''|$, which when combined with Eq.~\eqref{D_transform} implies
\begin{align}
-i\left(\sum_{h''} \eta_{h''}(p')\langle p',h''|D \right)  = -\frac{\partial}{\partial p^{\prime \mu}}  \left[\sum_{h''} \, p^{\prime \mu}  \eta_{h''}(p')\langle p',h''|\right]  +\Delta \sum_{h''} \eta_{h''}(p')\langle p',h''|,
\label{D_inter}
\end{align}
where we assume that dilational symmetry is unbroken, and hence: $D|0\rangle=0$. After acting with Eq.~\eqref{D_inter} on the state $| p,h\rangle$ and projecting on $\overline{\eta}_{h'}(p')$, one can use the orthogonality condition: $\overline{\eta}_{h'}(p')\eta_{h''}(p') = \delta_{h'h''}$, together with the on-shell state normalisation in Eq.~\eqref{norm}, to obtain the following expression for the matrix element of the dilation operator:
\begin{align}
\langle p',h'|D|p,h\rangle = i (2\pi)^{4}  \delta_{0}^{(+)}\!(p) \left[ -p^{\prime \mu} \, \overline{\eta}_{h'}(p')\frac{\partial  \eta_{h}}{\partial p^{\prime \mu}}(p') - \delta_{h'h} \, p^{\prime\mu}\frac{\partial}{\partial p^{\prime\mu}}    +  \delta_{h'h} \,(\Delta-4)  \right] \delta^{4}(p'-p).
\label{D_matrix1}
\end{align} 
For deriving GFF constraints it is simpler to work with the coordinates $(\bar{p},q)$. To do so, one can make use of the distributional identity in Eq.~\eqref{D_dist} of Appendix~\ref{appendix_a}, from which it follows: 
\begin{align}
\langle p',h'|D|p,h\rangle = -i (2\pi)^{4}  \delta_{0}^{(+)}\!(\bar{p}) \left[ \bar{p}^{\mu} \, \overline{\eta}_{h'}(\bar{p})\frac{\partial  \eta_{h}}{\partial \bar{p}^{ \mu}}(\bar{p}) + \delta_{h'h} \, \bar{p}^{\mu}\frac{\partial}{\partial q^{\mu}}    -  \delta_{h'h} \,(\Delta-1)  \right] \delta^{4}(q).
\label{D_eta}
\end{align}  
\ \\
\noindent
As in the case of the Poincar\'e charges~\cite{Cotogno:2019vjb, Lorce:2019sbq}, GFF constraints can be established by comparing Eq.~\eqref{D_eta} with the matrix element of $D$ obtained using the form factor decomposition in Eq.~\eqref{T_decomp}, together with the definition of the dilational current in Eq.~\eqref{D_current}. A rigorous expression for this matrix element is defined by:
\begin{align}
\langle p',h'|D|p,h\rangle &=  \lim_{\substack{d \rightarrow 0 \\ R \rightarrow \infty}} \int \ud^{4}x \ f_{d,R}(x)\, x_{\nu} \, e^{iq\cdot x} \,\langle p',h'|T^{0\nu}(0)|p,h\rangle \nonumber  \\
&= -i\lim_{\substack{d \rightarrow 0 \\ R \rightarrow \infty}} \frac{\partial \widetilde{f}_{d,R}(q)}{\partial q^{j}}\, \langle p',h'|T^{0j}(0)|p ,h\rangle = -i (2\pi)^{3} \partial_{j}\delta^{3}(\uvec{q})\, \langle p',h'|T^{0j}(0)|p ,h\rangle,
\label{D_eq1}
\end{align} 
where in order to ensure the convergence of the operator $D$ one integrates with a test function\footnote{See~\cite{Lowdon:2017idv} for an overview of these test function definitions and their motivation.} $f_{d,R}(x)=\alpha_{d}(x^{0})F_{R}(\uvec{x})$ that satisfies the conditions: $\int \ud x^{0} \, \alpha_{d}(x^{0}) =1$, $\lim_{d\rightarrow 0}\alpha_{d}(x^{0}) = \delta(x^{0})$, and $F_{R}(\uvec 0) = 1$, $\lim_{R\rightarrow \infty}F_{R}(\uvec{x}) = 1$, where $\widetilde{f}_{d,R}(q)$ is the Fourier transform, and $\partial_{j}$ indicates a derivative with respect to $q_{j}$. To evaluate this expression one therefore needs to understand how to simplify the final expression, which involves the product of a delta-derivative and a specific component of the EMT matrix element. These products were already encountered in~\cite{Cotogno:2019xcl} when deriving the form factor constraints due to Poincar\'e symmetry. Applying the distributional equality in Eq.~\eqref{D_dist2} of Appendix~\ref{appendix_a} to the coefficients of the GFFs appearing in the $(\mu=0,\nu=j)$ component of Eq.~\eqref{T_decomp}, together with the masslessness condition: $\bar{p}^{j}\bar{p}_{j} = -\bar{p}_{0}^{2}$ (for $q=0$), one obtains
\begin{align}
\langle p',h'|D|p,h\rangle &= -i (2\pi)^{4} \delta_{0}^{(+)}\!(\bar{p}) \bigg[- \bar{p}^{\mu}  \partial_{\mu} \!\left[\overline{\eta}_{h'}(p')\eta_{h}(p)\right]_{q=0} \delta^4(q) A(q^2)   + \delta_{h'h} \, \bar{p}^{\mu} \partial_{\mu}\delta^{4}(q) A(q^{2}) \nonumber \\
& \hspace{40mm}  - i\frac{\bar{p}_{j}}{\bar{p}^{0}} \, \overline{\eta}_{h'}(\bar{p})S^{0j}\eta_{h}(\bar{p})  \delta^{4}(q) G(q^{2})     \bigg].
\label{D_eq2}
\end{align} 
After using the conformal GPT constraint in Eq.~\eqref{conformal_GPT}, together with Eq.~\eqref{gpt_eq1} in Appendix~\ref{appendix_a}, the matrix elements can finally be written
 \begin{align}
\langle p',h'|D|p,h\rangle &= -i (2\pi)^{4} \delta_{0}^{(+)}\!(\bar{p}) \bigg[\bar{p}^{\mu} \, \overline{\eta}_{h'}(\bar{p})\frac{\partial  \eta_{h}}{\partial \bar{p}^{ \mu}}(\bar{p}) \,\delta^{4}(q) A(q^{2})   + \delta_{h'h} \, \bar{p}^{\mu} \partial_{\mu}\delta^{4}(q) A(q^{2}) \nonumber \\
& \hspace{45mm} - \delta_{hh'} (\Delta-1) \, \delta^{4}(q) G(q^{2})     \bigg].
\label{D_eq3}
\end{align}
Since Eqs.~\eqref{D_eta} and~\eqref{D_eq3} are different representations of the same matrix element, equating these expressions immediately leads to constraints on the GFFs, in particular:
\begin{align}
&A(q^{2}) \, \delta^{4}(q)  = \delta^{4}(q), \label{GFF1} \\
&A(q^{2}) \, \partial_{\mu}\delta^{4}(q)  =  \partial_{\mu}\delta^{4}(q),  \label{GFF2} \\
&G(q^{2}) \, \delta^{4}(q)  = \delta^{4}(q), \label{GFF3}
\end{align}
which is nothing more\footnote{Although the GFFs are in general distributions of $q$, one can interpret the values at $q=0$ using a limiting procedure~\cite{Lowdon:2017idv}. Eq.~\eqref{GFF2} follows trivially when $A(q^{2})$ is continuous at $q=0$ due to the $q^{2}$ dependence.} than the condition:
\begin{align}
A(0)=G(0)=1.
\end{align} 
That constraints are only imposed on $A(q^{2})$ and $G(q^{2})$ follows from the fact that the $x$-polynomiality order of the conserved currents has a direct bearing on whether the corresponding charges constrain certain GFFs. Since explicit factors of $x$ in the currents become $q$-derivatives on the level of the charges, as in Eq.~\eqref{D_eq1}, it is these derivatives that can remove powers of $q$ in Eq.~\eqref{T_decomp} and constrain the GFFs at $q=0$. Due to the structure of the dilational current [Eq.~\eqref{D_current}], the matrix elements of $D$ can therefore only potentially constrain the GFFs which have coefficients with at most one power of $q$. It is interesting to note that these are precisely the constraints obtained from imposing Poincar\'e symmetry alone~\cite{Cotogno:2019vjb,Lorce:2019sbq}.

\subsection{Special conformal covariance}
\label{SCT}

Now we perform an analogous procedure for SCTs. Using Eq.~\eqref{SC_transform} together with the fact that the special conformal symmetry is unbroken, and hence: $K^{\mu}|0\rangle=0$, one ends up with the following representation for the $K^{\mu}$ matrix elements:
\begin{align}
&\langle p',h'|K^{\mu}|p,h\rangle \nonumber = \\
&(2\pi)^{4} \delta_{0}^{(+)}\!(p) \bigg[ 2(\Delta-4)\overline{\eta}_{h'}(p')\frac{\partial \eta_{h}}{\partial p'_{\mu}}(p')\delta^{4}(p'-p) +  2(\Delta-4) \delta_{h'h}\frac{\partial}{\partial p'_{\mu}}\delta^{4}(p'-p)   \nonumber \\
& \quad\quad + \delta_{h'h}\left(p^{\prime \mu}\frac{\partial}{\partial p^{\prime \alpha}}\frac{\partial}{\partial p'_{\alpha}} -2p^{\prime \nu} \frac{\partial}{\partial p^{\prime \nu}}\frac{\partial}{\partial p'_{\mu}} \right)\delta^{4}(p'-p) \nonumber \\
& \quad\quad +\left(2p^{\prime \mu} \, \overline{\eta}_{h'}(p')\frac{\partial \eta_{h}}{\partial p^{\prime \alpha}}(p')\frac{\partial}{\partial p'_{\alpha}}   -2p^{\prime \nu} \,  \overline{\eta}_{h'}(p')\frac{\partial \eta_{h}}{\partial p^{\prime \nu}}(p')\frac{\partial}{\partial p'_{\mu}} -2p^{\prime \nu} \,  \overline{\eta}_{h'}(p')\frac{\partial \eta_{h}}{\partial p'_{\mu}}(p')\frac{\partial}{\partial p^{\prime \nu}} \right)\delta^{4}(p'-p)  \nonumber \\
& \quad\quad +\left(p^{\prime \mu} \, \overline{\eta}_{h'}(p')\frac{\partial^{2} \eta_{h}}{\partial p^{\prime \alpha} \partial p'_{\alpha}}(p') -2p^{\prime \nu} \, \overline{\eta}_{h'}(p')\frac{\partial^{2} \eta_{h}}{\partial p^{\prime \nu}\partial p'_{\mu}}(p') \right)\delta^{4}(p'-p)  \nonumber \\
& \quad\quad -2i  \overline{\eta}_{h'}(p') S^{\mu\nu} \frac{\partial \eta_{h}}{\partial p^{\prime \nu}}(p')\delta^{4}(p'-p)  -2i  \overline{\eta}_{h'}(p') S^{\mu\nu}\eta_{h}(p') \, \frac{\partial }{\partial p^{\prime \nu}}\delta^{4}(p'-p) \bigg].
\end{align} 
In contrast to the $D$ matrix elements, switching coordinates to $(\bar{p},q)$ in this expression is quite complicated due to the appearance of terms that involve more than one $p'$ derivative. Nevertheless, one can analyse the effect of this coordinate change on each of the non-trivial terms individually, which is summarised in Appendix~\ref{appendix_a}. After applying Eqs.~\eqref{dist_rel1}-\eqref{dist_rel4}, together with the GPT relations in Eqs.~\eqref{gpt_eq1}-\eqref{gpt_eq3}, one finally obtains:
\begin{align}
&\langle p',h'|K^{\mu}|p,h\rangle = \nonumber \\
&(2\pi)^{4} \delta_{0}^{(+)}\!(\bar{p}) \delta_{h'h} \left( \bar{p}^{\mu}\frac{\partial}{\partial q^{\nu}}\frac{\partial}{\partial q_{\nu}} -2\bar{p}^{\nu} \frac{\partial}{\partial q^{\nu}}\frac{\partial}{\partial q_{\mu}} \right)\delta^{4}(q) +(2\pi)^{4} \frac{\bar{p}^{\mu}}{2\bar{p}^{0}} \delta_{h'h}  \frac{\partial}{\partial \bar{p}^{0}}\delta_{0}^{(+)}\!(\bar{p}) \delta^{4}(q)     \nonumber \\ 
&+(2\pi)^{4} \delta_{0}^{(+)}\!(\bar{p}) \bigg[ -2(\Delta-1)\partial^{\mu} \!\left[\overline{\eta}_{h'}(p')\eta_{h}(p)\right]_{q=0}
 + \bar{p}^{\mu} \partial^{\alpha}\partial_{\alpha} \!\left[\overline{\eta}_{h'}(p')\eta_{h}(p)\right]_{q=0} \nonumber \\
& \quad\quad\quad\quad\quad\quad\quad\quad  -2\bar{p}^{\nu} \partial^{\mu}\partial_{\nu} \!\left[\overline{\eta}_{h'}(p')\eta_{h}(p)\right]_{q=0}   +2i \partial_{\nu} \!\left[\overline{\eta}_{h'}(p')S^{\mu\nu}\eta_{h}(p)\right]_{q=0} \bigg]\delta^{4}(q) \nonumber \\
&  + (2\pi)^{4} \delta_{0}^{(+)}\!(\bar{p}) \bigg[-2\bar{p}^{\mu}  \partial_{\nu} \!\left[\overline{\eta}_{h'}(p')\eta_{h}(p)\right]_{q=0}\frac{\partial}{\partial q_{\nu}}   +2\bar{p}^{\nu}  \partial_{\nu} \!\left[\overline{\eta}_{h'}(p')\eta_{h}(p)\right]_{q=0}\frac{\partial}{\partial q_{\mu}}  \nonumber \\
& \quad\quad\quad +2\bar{p}^{\nu} \partial^{\mu} \!\left[\overline{\eta}_{h'}(p')\eta_{h}(p)\right]_{q=0}\frac{\partial}{\partial q^{\nu}}  
 -2i  \overline{\eta}_{h'}(\bar{p}) S^{\mu\nu}\eta_{h}(\bar{p}) \, \frac{\partial }{\partial q^{\nu}}  +  2(\Delta-1) \delta_{h'h}\frac{\partial}{\partial q_{\mu}}   \bigg] \delta^{4}(q).  
 \label{K_eq1} 
\end{align} 
\ \\
\noindent 
Analogously to the case of dilations, one can now compare this expression to the matrix element derived from the form factor decomposition in Eq.~\eqref{T_decomp}. Using the modified form for the SCT current in Eq.~\eqref{K_current}, the matrix element of $K^{\mu}$ takes the form
\begin{align}
\langle p',h'|K^{\mu}|p,h\rangle &=  \lim_{\substack{d \rightarrow 0 \\ R \rightarrow \infty}} \int \ud^{4}x \, f_{d,R}(x)\, (2x^{\mu}x_{\alpha} - g^{\mu}_{\ \alpha}x^{2}) \, e^{iq\cdot x} \,\langle p',h'|T^{0\alpha}(0)|p,h\rangle \nonumber  \\
&=(2\pi)^{3} \left[ \partial_{j}\partial^{j}\delta^{3}(\uvec{q})\, \langle p',h'|T^{0\mu}(0)|p ,h\rangle - 2g^{\mu}_{\ k} \, \partial_{j}\partial^{k}\delta^{3}(\uvec{q})\, \langle p',h'|T^{0j}(0)|p ,h\rangle \right],
\label{K_eq2}
\end{align} 
where the inclusion of the test function $f_{d,R}(x)$ is again required in order ensure the convergence of the matrix element. The $x^{0}$-dependent terms are absent from this expression due to the definition of $f_{d,R}(x)$. In this case, one needs to understand the action of \textit{double} delta-derivative terms on the GFF components in order to evaluate Eq.~\eqref{K_eq2}. This complicated action is given by Eq.~\eqref{dd_dist} of Appendix~\ref{appendix_a}. Applying this relation, together with Eq.~\eqref{conformal_GPT}, one finally obtains the GFF representation
\begin{align}
&\langle p',h'|K^{\mu}|p,h\rangle =  \nonumber \\
&(2\pi)^{4} \delta_{0}^{(+)}\!(\bar{p}) \delta_{h'h} \left( \bar{p}^{\mu}\frac{\partial}{\partial q^{\nu}}\frac{\partial}{\partial q_{\nu}} -2\bar{p}^{\nu} \frac{\partial}{\partial q^{\nu}}\frac{\partial}{\partial q_{\mu}} \right)\delta^{4}(q)A(q^{2}) +(2\pi)^{4} \frac{\bar{p}^{\mu}}{2\bar{p}^{0}} \delta_{h'h}  \frac{\partial}{\partial \bar{p}^{0}}\delta_{0}^{(+)}\!(\bar{p}) \delta^{4}(q)A(q^{2})     \nonumber \\ 
&+(2\pi)^{4} \delta_{0}^{(+)}\!(\bar{p}) \bigg[ -2(\Delta-1)\partial^{\mu} \!\left[\overline{\eta}_{h'}(p')\eta_{h}(p)\right]_{q=0}G(q^{2})
 + \bar{p}^{\mu} \partial^{\alpha}\partial_{\alpha} \!\left[\overline{\eta}_{h'}(p')\eta_{h}(p)\right]_{q=0}A(q^{2}) \nonumber \\
& \quad\quad\quad\quad\quad\quad\quad\quad  -2\bar{p}^{\nu} \partial^{\mu}\partial_{\nu} \!\left[\overline{\eta}_{h'}(p')\eta_{h}(p)\right]_{q=0}A(q^{2})   +2i \partial_{\nu} \!\left[\overline{\eta}_{h'}(p')S^{\mu\nu}\eta_{h}(p)\right]_{q=0}G(q^{2}) \bigg]\delta^{4}(q) \nonumber \\
&  + (2\pi)^{4} \delta_{0}^{(+)}\!(\bar{p}) \bigg[-2\bar{p}^{\mu}  \partial_{\nu} \!\left[\overline{\eta}_{h'}(p')\eta_{h}(p)\right]_{q=0}A(q^{2})\frac{\partial}{\partial q_{\nu}}   +2\bar{p}^{\nu}  \partial_{\nu} \!\left[\overline{\eta}_{h'}(p')\eta_{h}(p)\right]_{q=0}A(q^{2})\frac{\partial}{\partial q_{\mu}}  \nonumber \\
&  +2\bar{p}^{\nu} \partial^{\mu} \!\left[\overline{\eta}_{h'}(p')\eta_{h}(p)\right]_{q=0}A(q^{2})\frac{\partial}{\partial q^{\nu}}  
 -2i  \overline{\eta}_{h'}(\bar{p}) S^{\mu\nu}\eta_{h}(\bar{p}) \, G(q^{2})\frac{\partial }{\partial q^{\nu}}  +  2(\Delta-1) \delta_{h'h}G(q^{2})\frac{\partial}{\partial q_{\mu}}   \bigg] \delta^{4}(q)   \nonumber \\
&-(2\pi)^{4}\delta_{0}^{(+)}\!(\bar{p}) \delta_{h'h} \frac{g^{\mu}_{\ 0}}{2\bar{p}^{0}}\left[A(q^{2})   - 2(\Delta-1)\, G(q^{2}) + 12 \,C(q^{2}) -4\Delta(\Delta-1)\, T(q^{2}) \right] \delta^{4}(q) \nonumber \\
&+(2\pi)^{4}\delta_{0}^{(+)}\!(\bar{p}) \delta_{h'h} \frac{g^{\mu}_{\ k}\bar{p}^{k}}{2(\bar{p}^{0})^{2}}\left[A(q^{2})   - 2(\Delta-1) \, G(q^{2}) + 12 \,C(q^{2}) -4\Delta(\Delta-1)\, T(q^{2}) \right] \delta^{4}(q).
\label{K_eq3}
\end{align} 
Equating the $K^{\mu}$ matrix element representations in Eqs.~\eqref{K_eq1} and Eq.~\eqref{K_eq3} one is then led to the following conditions:
\begin{align}
&A(q^{2}) \, \delta^{4}(q)  = \delta^{4}(q), \label{GFF1_K} \\
&A(q^{2}) \, \partial_{\mu}\delta^{4}(q)  =  \partial_{\mu}\delta^{4}(q),  \label{GFF2_K} \\
&A(q^{2}) \, \partial_{\mu}\partial_{\nu}\delta^{4}(q)  =  \partial_{\mu}\partial_{\nu}\delta^{4}(q),  \label{GFF4} \\
&G(q^{2}) \, \delta^{4}(q)  = \delta^{4}(q), \label{GFF3_K} \\
&G(q^{2}) \, \partial_{\mu}\delta^{4}(q)  =  \partial_{\mu}\delta^{4}(q),  \label{GFF5} \\
&\left[A(q^{2})   - 2(\Delta-1) \, G(q^{2}) + 12 \,C(q^{2}) -4\Delta(\Delta-1)\, T(q^{2}) \right] \delta^{4}(q) = 0, \label{GFF6} 
\end{align}
which ultimately imply the GFF constraints:
\begin{align}
&A(0)= G(0)=1, \label{K_1} \\
&(\partial_{\mu}\partial_{\nu}A)(0)=0, \label{K_2} \\
&A(0) - 2(\Delta-1) \, G(0) + 12 \,C(0) -4\Delta(\Delta-1)\, T(0)  = 0. \label{linear} 
\end{align}
So together with the constraints derived in Sec.~\ref{dilational} from dilational covariance, SCT covariance also introduces an additional constraint on the second derivative of $A(q^{2})$, and implies that all of the GFFs are in fact related at $q=0$ by a specific linear combination depending on the conformal dimension $\Delta$. In Sec.~\ref{enhance} it will be demonstrated that Eq.~\eqref{GFF6}, and hence Eq.~\eqref{linear}, are in fact further strengthened by the assumption made in this section that the SCT current has the form in Eq.~\eqref{K_current}.

\subsection{Conformal EMT trace constraints}
\label{trace_calc}

\subsubsection{General symmetric constraints}

The trace of the EMT plays an important role in the classification of CFTs. Since on-shell states are necessarily massless in any CFT, one can use the parametrisation in Eq.~\eqref{T_decomp} to determine the general action of $T^{\mu}_{\ \mu}$ on these states. In order to keep these calculations as general as possible we will first assume that the EMT is symmetric, but not necessarily traceless. After explicitly taking the trace in Eq.~\eqref{T_decomp} one obtains
\begin{align}
\langle p',h'|T^{\mu}_{\ \mu}(0)|p ,h \rangle &= \overline{\eta}_{h'}(p')\Big[- \tfrac{1}{2}q^{2}  A(q^{2}) + 2i \, \bar{p}_{\mu}S^{\mu\rho}q_{\rho} \, G(q^{2}) -6q^{2} \, C(q^{2})  \nonumber \\
& \hspace{30mm}   + 2 S^{\mu\alpha}S_{\mu}^{\ \beta}q_{\alpha}q_{\beta} \, T(q^{2})   \Big]\eta_{h}(p)  \, \delta_{0}^{(+)}\!(p')\,\delta_{0}^{(+)}\!(p).
\label{trace1}
\end{align}
To simplify this expression further one can make use of the conformal GPT relation in Eq.~\eqref{conformal_GPT}. The coefficient of $G(q^{2})$ becomes:
\begin{align}
2i \, \overline{\eta}_{h'}(p') S^{\mu\rho} \eta_{h}(p)  \bar{p}_{\mu}q_{\rho} &= -2i \, \overline{\eta}_{h'}(p') S^{\mu\rho} \eta_{h}(p)  p'_{\mu}p_{\rho} \nonumber \\
&= -2(p'\cdot p)(\Delta -1) \, \overline{\eta}_{h'}(p')\eta_{h}(p) =  q^{2}(\Delta -1) \, \overline{\eta}_{h'}(p')\eta_{h}(p),
\label{G_coeff}
\end{align} 
where the first equality follows from the anti-symmetry of $S^{\mu\rho}$. For the $T(q^{2})$ coefficient one instead obtains
\begin{align}
2 \, \overline{\eta}_{h'}(p')S^{\mu\alpha}S_{\mu}^{\ \beta}\eta_{h}(p) q_{\alpha}q_{\beta} = 2 \, \overline{\eta}_{h'}(p')S^{\mu\alpha}S_{\mu}^{\ \beta}\eta_{h}(p) ( p_{\alpha}p_{\beta} + p'_{\alpha}p'_{\beta} - p'_{\alpha}p_{\beta}- p_{\alpha}p'_{\beta}).
\label{T_mix}
\end{align}
Applying Eq.~\eqref{conformal_GPT} one can see that the first term in Eq.~\eqref{T_mix} vanishes since:
\begin{align}
\overline{\eta}_{h'}(p')S^{\mu\alpha}S_{\mu}^{\ \beta}\eta_{h}(p)p_{\alpha}p_{\beta} = -i(\Delta -1)\overline{\eta}_{h'}(p')S^{\mu\alpha}\eta_{h}(p)p_{\alpha}p_{\mu} = 0.
\end{align} 
Taking the dual of the second term it follows that this term similarly vanishes. For the third term, one instead finds that:
\begin{align}
-2 \, \overline{\eta}_{h'}(p')S^{\mu\alpha}S_{\mu}^{\ \beta}\eta_{h}(p)p'_{\alpha}p_{\beta} &= 2i(\Delta-1) \, \overline{\eta}_{h'}(p')S^{\mu\alpha}\eta_{h}(p)p'_{\alpha}p_{\mu} \nonumber \\
&= -2(p'\cdot p)(\Delta-1)^{2} \, \overline{\eta}_{h'}(p')\eta_{h}(p) = q^{2}(\Delta-1)^{2} \, \overline{\eta}_{h'}(p')\eta_{h}(p).
\end{align}
Using the fact that the Lorentz generators in any field representation satisfy 
\begin{align}
\left[S^{\mu\alpha},S^{\nu\beta}\right] = i(g^{\mu\beta}S^{\alpha\nu} + g^{\alpha\nu}S^{\mu\beta} -g^{\mu\nu}S^{\alpha\beta} -g^{\alpha\beta}S^{\mu\nu}),
\end{align}
and hence: $S^{\mu\alpha}S_{\mu}^{\ \beta} = S_{\mu}^{\ \beta}S^{\mu\alpha} -2i\, S^{\alpha\beta}$, the last term can then be written
\begin{align}
-2 \, \overline{\eta}_{h'}(p')S^{\mu\alpha}S_{\mu}^{\ \beta}\eta_{h}(p)p_{\alpha}p'_{\beta} &= -2 \, \overline{\eta}_{h'}(p')\left[S_{\mu}^{\ \beta}S^{\mu\alpha} -2i\, S^{\alpha\beta} \right]\eta_{h}(p)p_{\alpha}p'_{\beta} \nonumber \\
&= -2 \, \overline{\eta}_{h'}(p')\left[-iS_{\mu}^{\ \beta}p^{\mu}p'_{\beta}(\Delta-1) +2(\Delta-1)(p'\cdot p)  \right]\eta_{h}(p) \nonumber \\
&= q^{2}\left[(\Delta-1)^{2} +2(\Delta-1) \right] \, \overline{\eta}_{h'}(p')\eta_{h}(p).
\end{align} 
After combining all of these results, the $T(q^{2})$ coefficient takes the form
\begin{align}
2 \, \overline{\eta}_{h'}(p')S^{\mu\alpha}S_{\mu}^{\ \beta}\eta_{h}(p) q_{\alpha}q_{\beta} &= \left\{q^{2}\left[(\Delta-1)^{2} +2(\Delta-1) \right] + q^{2}(\Delta-1)^{2} \right\} \, \overline{\eta}_{h'}(p')\eta_{h}(p) \nonumber \\
&= 2\Delta(\Delta-1)q^{2} \, \overline{\eta}_{h'}(p')\eta_{h}(p).
\label{T_coeff}
\end{align}
Inserting Eqs.~\eqref{G_coeff} and~\eqref{T_coeff} into Eq.~\eqref{trace1}, one finally obtains the following expression for the trace matrix element:
\begin{align}
\langle p',h'|T^{\mu}_{\ \mu}(0)|p ,h \rangle &= - \frac{1}{2}q^{2}\Big[  A(q^{2}) - 2(\Delta -1) \, G(q^{2}) +12 \, C(q^{2})  \nonumber \\
&\hspace{30mm} - 4\Delta(\Delta-1) \, T(q^{2}) \Big]\overline{\eta}_{h'}(p')\eta_{h}(p)  \, \delta_{0}^{(+)}\!(p')\,\delta_{0}^{(+)}\!(p).
\label{trace2}
\end{align}
Eq.~\eqref{trace2} demonstrates an important structural feature of CFTs: although the GFFs in Eq.~\eqref{T_decomp} have coefficients with different $q$ dependencies, taking the trace results in an expression with an overall $q^{2}$ coefficient. The relevance of this feature will be discussed in more detail in Sec.~\ref{massless_part}.

\subsubsection{Enhancement of the SCT constraints}
\label{enhance}

As established in Sec.~\ref{SCT}, the requirement of SCT covariance implies that the corresponding GFFs are linearly related to one another at $q=0$. In deriving this constraint we implicitly assumed that the SCT current has the form in Eq.~\eqref{K_current}, and hence the EMT is both symmetric and traceless. We will now demonstrate that the tracelessness of the EMT leads to a strengthening of the constraint in Eq.~\eqref{linear}. By demanding that $T^{\mu}_{\ \mu}(x)=0$, it follows from Eq.~\eqref{trace2} that
\begin{align}
q^{2}\Big[  A(q^{2}) - 2(\Delta -1) \, G(q^{2}) +12 \, C(q^{2})  - 4\Delta(\Delta-1) \, T(q^{2}) \Big]  \, \delta_{0}^{(+)}\!(p')\,\delta_{0}^{(+)}\!(p)=0.
\end{align}        
Switching to the variables $(\bar{p},q)$, this implies the distributional equality 
\begin{align}
\left[\frac{(\bar{\uvec{p}}^{2} -\tfrac{1}{4}\uvec{q}^{2})}{|\bar{\uvec{p}}+ \tfrac{1}{2}\uvec{q}||\bar{\uvec{p}}- \tfrac{1}{2}\uvec{q}|} -1 \right] \Big[  A(q^{2}) - 2(\Delta -1) \, G(q^{2}) +12 \, C(q^{2})  - 4\Delta(\Delta-1) \, T(q^{2}) \Big]   =0.
\end{align} 
Since the coefficient of this expression vanishes only at $\uvec{q}=\uvec{0}$, the linear combination of form factors has the general solution 
\begin{align}
A(q^{2}) - 2(\Delta -1) \, G(q^{2}) +12 \, C(q^{2}) - 4\Delta(\Delta-1) \, T(q^{2}) = \mathcal{C} \, \delta^{3}(\uvec{q}),
\end{align}
where $\mathcal{C}$ is some arbitrary distribution in $q^{0}$. However, in order for this expression to be compatible with Eq.~\eqref{GFF6} it must be the case that $\mathcal{C}\equiv  0$, otherwise one would end up with the ill-defined product: $\delta^{3}(\uvec{q})\delta^{3}(\uvec{q})$. So by explicitly taking into account the tracelessness of the EMT, this implies that Eq.~\eqref{linear} is in fact a realisation of the more general condition:
\begin{align}
A(q^{2}) - 2(\Delta -1) \, G(q^{2}) +12 \, C(q^{2}) - 4\Delta(\Delta-1) \, T(q^{2}) = 0.
\label{DNA_eq}
\end{align}
Eq.~\eqref{DNA_eq} together with Eqs.~\eqref{K_1} and \eqref{K_2} collectively summarise the constraints imposed on the EMT matrix elements by Poincar\'e and conformal symmetry. In the next section we will demonstrate, using explicit CFT examples, that these constraints are sufficient to completely specify the form of these matrix elements. This is not necessarily surprising since it is well known that the structural form of correlation functions in CFTs are fixed by the overall symmetry, and in particular, the closely related EMT three-point functions are determined by the conformal Ward identities, as discussed in Sec.~\ref{intro}. \\

\noindent
Before outlining specific examples in the next section, we first draw attention to the observation made in Sec.~\ref{dilational} that the $x$-polynomiality order of the conserved currents determines whether certain GFFs can be constrained by the corresponding symmetry. This explains why the dilational charge matrix elements only result in constraints on $A(q^{2})$ and $G(q^{2})$, whereas the SCT charge can also constrain $C(q^{2})$ and $T(q^{2})$, both of which have coefficients involving two powers of $q$. Since conformal symmetry is expected to completely constrain the structure of any matrix element, and the conformal currents involve at most two powers of $x$, this implies that local and covariant EMT matrix elements in CFTs can only ever contain form factors that have coefficients with at most \textit{two} powers of $q$, otherwise the conformal symmetry would not be sufficient to fully constrain the matrix elements. As pointed out at the end of Sec.~\ref{grav_FF}, in local QFTs it turns out that the masslessness of the states alone is actually sufficient to guarantee that this is indeed the case. This further emphasises the close connection between the presence of massless particles and the existence of conformal symmetry. We will explore this connection in more detail in Sec.~\ref{massless_part}.

\subsection{Explicit CFT examples}
\label{examples}

\subsubsection{Free massless scalar theory}

The simplest example of a unitary CFT is that of a free massless scalar field $\phi$. The states have $h=0$, and the Lorentz generators appearing in Eq.~\eqref{T_decomp} are trivial, hence the only form factors that can exist are $A_{\phi}(q^{2})$ and $C_{\phi}(q^{2})$. Eq.~\eqref{DNA_eq} therefore takes the form:
\begin{align}
A_{\phi}(q^{2}) + 12 \, C_{\phi}(q^{2})= 0.
\label{scalar_DNA}
\end{align} 
Due to the absence of interactions, and the fact that $q^{2}$ is the only dimensionful parameter in the theory, it follows that $A_{\phi}(q^{2})$ must be constant. Combining this with the constraint in Eq.~\eqref{K_1}, Eq.~\eqref{scalar_DNA} immediately implies\footnote{This result coincides with that found in~\cite{Hudson:2017xug}, where the authors instead use the form factor $D(q^{2})= 4 C(q^{2})$.}  
\begin{align}
C_{\phi}(q^{2}) = -\tfrac{1}{12}.
\label{C_scalar}
\end{align} 
So the conformal symmetry, together with the masslessness of the states, completely fixes the matrix elements of the symmetric-traceless EMT of the scalar field.

\subsubsection{Free massless fermion theory}

Another simple example of a unitary CFT is the free massless fermion $\psi$. Since the parametrisation in Eq.~\eqref{T_decomp} assumes the EMT is both $\mathsf{P}$ and $\mathsf{T}$ invariant, for consistency $\psi$ must therefore be in the Dirac representation\footnote{We assumed for simplicity in Eq.~\eqref{T_decomp} that the EMT is invariant under discrete symmetries. This requirement could of course be loosened, which would result in more potential form factor structures, and enable one to analyse CFTs with fields in non $\mathsf{P}$ or $\mathsf{T}$-symmetric representations, such as Weyl fermions.}. In this case the states can have $h= \pm\tfrac{1}{2}$, and the only independent form factors are: $A_{\psi}(q^{2})$, $G_{\psi}(q^{2})$, and $C_{\psi}(q^{2})$, hence Eq.~\eqref{DNA_eq} takes the form:
\begin{align}
A_{\psi}(q^{2}) - 2(\Delta -1) \, G_{\psi}(q^{2}) + 12 \, C_{\psi}(q^{2})= 0.
\label{fermion_DNA}
\end{align} 
As in the scalar case: $A_{\psi}(q^{2})=1$, but also the absence of interactions and Eq.~\eqref{K_1} implies: $G_{\psi}(q^{2})=1$. Combining these conditions with Eq.~\eqref{fermion_DNA}, it follows that:
\begin{align}
C_{\psi}(q^{2}) = -\tfrac{1}{12}(3 - 2\Delta) =0,
\label{C_fermionc}
\end{align} 
where the last equality is due to the fact that the corresponding GPTs in Eq.~\eqref{T_decomp} have $\Delta=\tfrac{3}{2}$. So although the EMT matrix elements for $h=\pm\tfrac{1}{2}$ states can potentially have more covariant structures than those with $h=0$, the conformal symmetry and masslessness of the states is still sufficient to completely fix the form of the EMT matrix elements.

\subsubsection{Massless theories with $|h| \geq 1$ states}
\label{higher_h}

As already discussed in Sec.~\ref{grav_FF}, by virtue of the Weinberg-Witten Theorem, Eq.~\eqref{T_decomp} can no longer hold for arbitrary states with $|h|>1$. This does not mean that no parametrisation exists, only that for a unitary theory this parametrisation cannot be both local and covariant~\cite{Lopuszanski:1983zx}. It is interesting to note that this theorem does not explicitly rule out the possibility that Eq.~\eqref{T_decomp} is satisfied for theories containing massless states with $h=\pm 1$. A simple example is the theory of free photons. This CFT is constructed from the anti-symmetric tensor field $F_{\mu\nu}$, which by virtue of Eq.~\eqref{conformal_field} satisfies the free Maxwell equations. Due to the Poincar\'e Lemma it follows that $F_{\mu\nu}$ cannot be fundamental, but instead must involve the derivative of another field: $F_{\mu\nu} = \partial_{\mu}A_{\nu} - \partial_{\nu}A_{\mu}$. By treating the massless field $A_{\mu}$ to be fundamental, the resulting theory is invariant under gauge symmetry. However, an important consequence of this gauge symmetry is that it prevents $A_{\mu}$ from being \textit{both} local and Poincar\'e covariant~\cite{Strocchi13}. The corresponding EMT matrix elements of the free photon states must therefore necessarily either violate locality or covariance, and hence the parametrisation in Eq.~\eqref{T_decomp} cannot hold in general. \\

\noindent
As is well known, in order to make sense of gauge theories one must either permit non-local and non-covariant fields, such as in Coulomb gauge, or perform a gauge-fixing that preserves locality and covariance, but allows for the possibility of states with non-positive norm, like Gupta-Bleuler quantisation~\cite{Strocchi13}. In the latter case, it turns out that one can in fact recover a manifestly local and covariant EMT decomposition which coincides with Eq.~\eqref{T_decomp} for the \textit{physical} photon states. The difference to the lower helicity examples is that although $A_{A}(q^{2})$, $G_{A}(q^{2})$, $C_{A}(q^{2})$, and $T_{A}(q^{2})$ are actually non-vanishing, the tracelessness of the EMT does not result in the constraint in Eq.~\eqref{DNA_eq}, since $A_{\mu}$ is not a conformal field. In general, for massless fields that create states with higher helicity ($|h|>1$) the Poincar\'e Lemma equally applies, and hence similarly forces the introduction of non-covariant gauge-dependent fields~\cite{Lopuszanski:1983zx}. In this sense, the existence of massless particles with $|h| \geq 1$ is intimately connected with the presence of gauge symmetry. Before concluding, we note that one could also equally perform a local and covariant gauge-fixing procedure in theories with massless $|h|>1$ states, such as the graviton. Due to dimensional arguments one would equally expect the EMT matrix elements of the physical modes to have the structure of Eq.~\eqref{T_decomp}, but we leave an investigation of these issues to a future work.

\section{Free massless particle constraints}
\label{massless_part}

Although the analysis in Sec.~\ref{matrix_constr} implicitly assumes that the GFFs, and the constraints imposed upon them, correspond to those of the modified current $T_{\!(\text{ST})}^{\mu\nu}$, the decomposition in Eq.~\eqref{T_decomp} equally holds for any choice of symmetric EMT, $T^{\mu\nu}_{\!(\text{S})}$. In what follows, we will use this expression to further explore the conditions under which unitary conformality holds, focussing in particular on the case of free theories with massless particle states. \\

\noindent
In general, given a QFT that is unitary, local, and Poincar\'e covariant, it follows from Eq.~\eqref{trace1} that the massless one-particle trace matrix elements of any symmetric EMT $T^{\mu\nu}_{\!(\text{S})}$ have the general form:
\begin{align}
\langle p',h'|T^{\mu}_{\!(\text{S})\mu}(0)|p ,h \rangle = -q^{2}F(q^{2})\, \overline{\eta}_{h'}(p')\eta_{h}(p)  \, \delta_{0}^{(+)}\!(p')\,\delta_{0}^{(+)}\!(p), 
\label{trace_cond}
\end{align}
where $F(q^{2})$ is a local form factor\footnote{By \textit{local} we mean that the form factor arises from the matrix elements of a local EMT current. In particular, this implies that certain types of components are excluded from appearing in the form factors, such as inverse powers of momentum. See~\cite{Lopuszanski:1983zx} for an in-depth discussion of this issue.}. For states with $h=0$ only the form factors $A(q^{2})$ and $C(q^{2})$ can appear, and so Eq.~\eqref{trace_cond} holds with
\begin{align}
F_{h=0}(q^{2}) = \frac{1}{2}\big[ A(q^{2})  +12 \, C(q^{2}) \big].
\end{align}
For $h \neq 0$ things are more complicated since the other form factors $G(q^{2})$ and $T(q^{2})$ can also potentially exist, and so one necessarily needs to understand how the Lorentz generators $S^{\mu\nu}$ act on the massless GPTs $\eta_{h}(p)$ in order to evaluate the trace of the EMT matrix elements. It turns out though that for free \textit{irreducible}\footnote{The irreducible massless fields are precisely those with Lorentz representations $(m,0)$, $(0,n)$, or their direct sums.} massless fields $\Phi(x)$, the fields satisfy the general relation: 
\begin{align}
\mathcal{C}_{\Phi} \, \partial^{\mu}\Phi(x) = i S^{\mu\nu}\partial_{\nu}\Phi(x),
\end{align}
and hence the corresponding GPTs obey the condition:
\begin{align}
\mathcal{C}_{\Phi} \, p^{\mu}\eta_{h}(p) = ip_{\nu} S^{\mu\nu}\eta_{h}(p),
\label{general_GPT}
\end{align}
where $\mathcal{C}_{\Phi}$ is some constant depending on the representation of the field. Since Eq.~\eqref{general_GPT} has the same form as Eq.~\eqref{conformal_GPT}, and Eq.~\eqref{T_decomp} only involves the GPTs of irreducible fields, as discussed in Sec.~\ref{grav_FF}, one can perform an identical calculation to that in Sec.~\ref{trace_calc}, similarly arriving at an expression for the one-particle matrix elements of $T^{\mu}_{\!(\text{S})\mu}$ with the structure of Eq.~\eqref{trace_cond}. \\

\noindent 
Now that we have established that Eq.~\eqref{trace_cond} is a generic feature of \textit{any} unitary, local, Poincar\'e covariant QFT with massless on-shell states\footnote{For massive states, one can immediately see that Eq.~\eqref{trace_cond} is violated, since the leading order component in the form factor expansion $\bar{p}^{\{\mu}\bar{p}^{\nu\}}A(q^{2})$ introduces an additional term $2M^{2}A(q^{2})$ to the trace matrix element, which cannot be written in form $q^{2}F(q^{2})$ since this would require explicit inverse powers of momentum, violating locality~\cite{Lopuszanski:1983zx}.}, one can explore the implications of this relation for free theories. Firstly, it follows from Eq.~\eqref{trace_cond} that:
\begin{align}
\langle p',h'|T^{\mu}_{\!(\text{S})\mu}(x)|p ,h \rangle =  \partial^{2} \left[e^{iq\cdot x}F(q^{2})\, \overline{\eta}_{h'}(p')\eta_{h}(p)  \, \delta_{0}^{(+)}\!(p')\,\delta_{0}^{(+)}\!(p)\right],
\label{invert}
\end{align}
which upon inversion gives:
\begin{align}
\langle p',h'|(\partial^{2})^{-1} T^{\mu}_{\!(\text{S})\mu}(x)|p ,h \rangle=  e^{iq\cdot x}F(q^{2})\, \overline{\eta}_{h'}(p')\eta_{h}(p)  \, \delta_{0}^{(+)}\!(p')\,\delta_{0}^{(+)}\!(p).  
\label{invert2}
\end{align}
The appearance of additional terms from the inversion involving first and zero-order polynomials in $x$ is ruled out by the translational covariance of the EMT. Since by definition $F(q^{2})$ contains no non-local contributions, Eq.~\eqref{invert2} therefore implies that $(\partial^{2})^{-1}T^{\mu}_{\!(\text{S})\mu}(x)$ acts like a (translationally covariant) local operator on the one-particle states. For free theories with multiple particle species this argument extends to the full space of one-particle states $\mathcal{H}_{1}$ because the EMT is diagonal in the different fields, due to the absence of interactions, and hence the action of $(\partial^{2})^{-1}T^{\mu}_{\!(\text{S})\mu}(x)$ on $\mathcal{H}_{1}$ corresponds to the sum of the local operators that exist for each state. For multi-particle states this argument can also be naturally extended, since the action of translations $U(a)=e^{iP\cdot a}$ on these states takes the form: 
\begin{align}
U(a) |p_{1},h_{1}; p_{2},h_{2}; \, \cdots\,  p_{n},h_{n} \rangle = U(a)|p_{1},h_{1}\rangle \, U(a)|p_{2},h_{2}\rangle \cdots U(a)|p_{n},h_{n}\rangle,
\label{multi}
\end{align}
where the inner product of $|p_{1},h_{1}; p_{2},h_{2}; \, \cdots\,  p_{n},h_{n} \rangle$ is constructed by taking the weighted sum of all possible products of one-particle inner products, with weight $+1$ or $-1$ depending on the helicity of the states~\cite{Haag:1992hx}. Therefore, by acting with $\left.\frac{d}{d a}\right|_{a=0}$ on Eq.~\eqref{multi} it is clear that the multi-particle matrix elements of $P^{\mu}$, and hence the EMT, are fixed by the corresponding one-particle matrix elements. Since for any free theory the full space of states $\mathcal{H}$ is spanned entirely by multi-particle (Fock) states, the previous arguments imply that the EMT trace has the form: $T^{\mu}_{\!(\text{S})  \mu}(x) = \partial^{2}L(x)$ on $\mathcal{H}$, with $L(x)$ some local scalar operator. This is precisely the condition for a unitary QFT to be conformal, as outlined in Sec.~\ref{conf_FC}, and hence one is led to the following conclusion:
\begin{displayquote}
\textit{Any unitary, local, and Poincar\'e covariant QFT, comprised solely of massless free fields, is conformal.}  
\end{displayquote}
This demonstrates that the presence of massless particles is not only relevant for the general structural properties of local QFTs, but also plays a significant role in determining the allowed symmetries. From a similar perspective, in the work of Buchholz and Fredenhagen~\cite{Buchholz:1976hz} it was proven for massless scalar fields that dilational covariance is sufficient to imply that the fields must be free. This was later extended by Weinberg to (irreducible) massless fields with higher values of absolute helicity $|h|$~\cite{Weinberg:2012cd}. Since any field in a conformal theory is dilationally covariant, one can therefore combine the results of Buchholz, Fredenhagen and Weinberg with the conclusion derived above, which immediately implies the stronger result: 
\begin{displayquote}
\centering
\textit{If $\Phi(x)$ is an irreducible massless field in a unitary, local, Poincar\'e covariant QFT: \\
$\Phi(x)$ is free \ $\Longleftrightarrow$ \ $\Phi(x)$ is dilationally covariant}  
\end{displayquote}
This means that dilational covariance is not only a sufficient condition for a massless field to be free, but is also necessary. In other words, the space of dilationally symmetric theories with massless particles is identical to the space of massless free theories. This runs in stark contrast to the case of massive particles, where spacetime symmetries alone are not sufficient to constrain whether or not the theories possess interactions.

\section{Conclusions}
\label{concl}

It is well known that conformal symmetry imposes significant constraints on the structure of conformal field theories (CFTs), in particular the correlation functions. In this work we investigate four-dimensional unitary, local, and Poincar\'e covariant CFTs, focussing on the analytic properties of the energy-momentum tensor (EMT) and the corresponding on-shell matrix elements. By adopting a parametrisation in terms of covariant multipoles of the Lorentz generators, we establish a local and covariant form factor decomposition of these matrix elements for states of general helicity. Using this decomposition, we derive the explicit constraints imposed on the form factors due to conformal symmetry and the trace properties of the EMT, and illustrate with specific CFT examples that they uniquely fix the form of the matrix elements. We also use this decomposition to gain new insights into the constraints imposed by the existence of massless particles, demonstrating in particular that massless free theories are conformal. Besides the applications outlined in this work, this matrix element decomposition could also be used to shed light on other aspects of massless QFTs, such as model-independent constraints like the averaged null energy condition~\cite{Klinkhammer:1991ki,Faulkner:2016mzt,Hartman:2016lgu}, and conformal collider bounds~\cite{Hofman:2008ar}. Although we have focussed here on the on-shell matrix elements of the EMT, which are projections of the subset of three-point functions involving the EMT, the same covariant multipole approach is equally applicable to more general CFT correlation functions, and could enable helicity-universal representations of these objects to be similarly derived.

\section*{Acknowledgements}

This work was supported by the Agence Nationale de la Recherche under the Projects No. ANR-18-ERC1-0002 and No. ANR-16-CE31-0019. The authors would like to thank Peter Schweitzer, Guillaume Bossard, and Christoph Kopper for useful discussions. The authors are also grateful to the referee for their constructive comments.

\appendix

\section{Distributional relations}
\label{appendix_a}

To derive the various constraints in Sec.~\ref{matrix_constr} it is necessary to change coordinates from $(p',p)$ to $(\bar{p},q)$. In order to do so, one makes use of the following relations:
\begin{align}
&\delta_{0}^{(+)}\!(p) \, p^{\prime\mu}\frac{\partial}{\partial p^{\prime\mu}} \delta^{4}(p'-p) = -3 \, \delta_{0}^{(+)}\!(\bar{p})\, \delta^{4}(q) + \delta_{0}^{(+)}\!(\bar{p}) \, \bar{p}^{\mu}\frac{\partial}{\partial q^{\mu}} \delta^{4}(q), \label{D_dist} \\
&\delta_{0}^{(+)}\!(p) \, \frac{\partial}{\partial p'_{\mu}} \delta^{4}(p'-p) = \frac{\bar{p}^{\mu}}{2\bar{p}^{0}}\frac{\partial}{\partial\bar{p}^{0}} \delta_{0}^{(+)}\!(\bar{p}) \, \delta^{4}(q) + \delta_{0}^{(+)}\!(\bar{p}) \, \frac{\partial}{\partial q_{\mu}}\delta^{4}(q), \label{dist_rel1} \\
&\delta_{0}^{(+)}\!(p) \, p'_{\alpha} \overline{\eta}_{h'}(p')\frac{\partial \eta_{h}}{\partial p^{\prime \nu}}(p') \frac{\partial}{\partial p'_{\mu}} \delta^{4}(p'-p) = \nonumber \\
& \hspace{10mm} \overline{\eta}_{h'}(\bar{p})\frac{\partial \eta_{h}}{\partial \bar{p}^{\nu}}(\bar{p}) \bigg[ \frac{\bar{p}^{\mu}\bar{p}_{\alpha}}{2\bar{p}^{0}}\frac{\partial}{\partial\bar{p}^{0}} \delta_{0}^{(+)}\!(\bar{p}) \, \delta^{4}(q) + \bar{p}_{\alpha} \, \delta_{0}^{(+)}\!(\bar{p}) \, \frac{\partial}{\partial q_{\mu}}\delta^{4}(q)  -\frac{1}{2}g^{\mu}_{\ \alpha}\delta_{0}^{(+)}\!(\bar{p}) \, \delta^{4}(q)  \bigg]  \nonumber \\
&\hspace{10mm} -\frac{1}{2}\bar{p}_{\alpha} \left[\frac{\partial \overline{\eta}_{h'}}{\partial \bar{p}_{\mu}}(\bar{p}) \frac{\partial \eta_{h}}{\partial \bar{p}^{\nu}}(\bar{p}) + \overline{\eta}_{h'}(\bar{p})\frac{\partial^{2} \eta_{h}}{\partial \bar{p}^{\nu} \partial \bar{p}_{\mu}}(\bar{p})    \right]\delta_{0}^{(+)}\!(\bar{p})\delta^{4}(q), \label{dist_rel2} \\
&\delta_{0}^{(+)}\!(p) \, \overline{\eta}_{h'}(p') S^{\mu\nu}\eta_{h}(p')\frac{\partial}{\partial p^{\prime \nu}} \delta^{4}(p'-p) = \nonumber \\
&\hspace{10mm}  \overline{\eta}_{h'}(\bar{p}) S^{\mu\nu}\eta_{h}(\bar{p}) \bigg[ \frac{\bar{p}_{\nu}}{2\bar{p}^{0}}\frac{\partial}{\partial\bar{p}^{0}} \delta_{0}^{(+)}\!(\bar{p}) \, \delta^{4}(q) + \delta_{0}^{(+)}\!(\bar{p}) \, \frac{\partial}{\partial q^{\nu}}\delta^{4}(q) \bigg] \nonumber \\
&\hspace{10mm} - \frac{1}{2}\left[   \frac{\partial \overline{\eta}_{h'}}{\partial \bar{p}^{\nu}}(\bar{p})S^{\mu\nu} \eta_{h}(\bar{p})  + \overline{\eta}_{h'}(\bar{p})S^{\mu\nu}\frac{\partial \eta_{h}}{\partial \bar{p}^{\nu}}(\bar{p})  \right]\delta_{0}^{(+)}\!(\bar{p})  \delta^{4}(q),  \label{dist_rel3} \\
&\delta_{0}^{(+)}\!(p) \left(p^{\prime \mu}\frac{\partial}{\partial p^{\prime \nu}}\frac{\partial}{\partial p'_{\nu}} -2p^{\prime \nu} \frac{\partial}{\partial p^{\prime \nu}}\frac{\partial}{\partial p'_{\mu}} \right)\delta^{4}(p'-p)=  \nonumber \\
& \hspace{5mm} \delta_{0}^{(+)}\!(\bar{p}) \left( \bar{p}^{\mu}\frac{\partial}{\partial q^{\nu}}\frac{\partial}{\partial q_{\nu}} -2\bar{p}^{\nu} \frac{\partial}{\partial q^{\nu}}\frac{\partial}{\partial q_{\mu}} \right)\delta^{4}(q) +  \frac{7\bar{p}^{\mu}}{2\bar{p}^{0}}\frac{\partial}{\partial\bar{p}^{0}} \delta_{0}^{(+)}\!(\bar{p}) \, \delta^{4}(q) + 6\, \delta_{0}^{(+)}\!(\bar{p}) \, \frac{\partial}{\partial q_{\mu}}\delta^{4}(q), \label{dist_rel4} \\
& \overline{\eta}_{h'}(\bar{p})\frac{\partial \eta_{h}}{\partial \bar{p}_{\mu}}(\bar{p}) = -\frac{\partial}{\partial q_{\mu}} \!\left[\overline{\eta}_{h'}(p')\eta_{h}(p)\right]_{q=0}, \label{gpt_eq1} \\
&\frac{\partial \overline{\eta}_{h'}}{\partial \bar{p}_{\mu}}(\bar{p})\frac{\partial \eta_{h}}{\partial \bar{p}^{\nu}}(\bar{p}) + \frac{\partial \overline{\eta}_{h'}}{\partial \bar{p}^{\nu}}(\bar{p})\frac{\partial \eta_{h}}{\partial \bar{p}_{\mu}}(\bar{p})  = -2 \frac{\partial}{\partial q_{\mu}}\frac{\partial}{\partial q^{\nu}} \!\left[\overline{\eta}_{h'}(p')\eta_{h}(p)\right]_{q=0}, \label{gpt_eq2} \\
&  \frac{\partial \overline{\eta}_{h'}}{\partial \bar{p}^{\nu}}(\bar{p}) S^{\mu\nu} \eta_{h}(\bar{p}) - \overline{\eta}_{h'}(\bar{p}) S^{\mu\nu} \frac{\partial \eta_{h}}{\partial \bar{p}^{\nu}}(\bar{p}) = 2 \frac{\partial}{\partial q^{\nu}} \!\left[\overline{\eta}_{h'}(p')S^{\mu\nu}\eta_{h}(p)\right]_{q=0}.  \label{gpt_eq3}
\end{align}
As opposed Eqs.~\eqref{gpt_eq1}-\eqref{gpt_eq3}, which follow immediately from the definition of the variables $(\bar{p},q)$, Eqs.~\eqref{D_dist}-\eqref{dist_rel4} are equalities between distributions, and so to derive them one needs to explicitly determine their action on test functions. Since the derivation of these various relations is rather similar, we will not repeat them all here but instead focus on proving Eq.~\eqref{dist_rel1}. Integrating this expression with the test function $f(p',p) = \bar{f}(\bar{p},q)$, and performing a change of variable, one obtains
\begin{align*}
\int d^{4}p' \, d^{4}p \ \delta_{0}^{(+)}\!(p) \, \frac{\partial}{\partial p'_{\mu}} \delta^{4}(p'-p) \, f(p',p) &= 2\pi \!\int d^{4}\bar{p} \, d^{4}q \ \frac{\delta\!\left(\bar{p}^{0} - \tfrac{1}{2}q^{0} - \sqrt{(\bar{\uvec{p}} -\tfrac{1}{2}\uvec{q})^{2}} \right)}{2\sqrt{(\bar{\uvec{p}} -\tfrac{1}{2}\uvec{q})^{2}}} \, \frac{\partial}{\partial q_{\mu}} \delta^{4}(q) \, \bar{f}(\bar{p},q)  \\
&= -2\pi  \int d^{3}\bar{\uvec{p}} \, d^{4}q  \, \frac{\partial}{\partial q_{\mu}} \left[ \frac{   \bar{f}\!\left(\bar{p}^{0}_{\star}, \bar{\uvec{p}}, q \right) }{2\sqrt{(\bar{\uvec{p}} -\tfrac{1}{2}\uvec{q})^{2}}} \right]\!  \delta^{4}(q),
\end{align*}
where $\bar{p}^{0}_{\star}=\tfrac{1}{2}q^{0} + \sqrt{(\bar{\uvec{p}} -\tfrac{1}{2}\uvec{q})^{2}}$. Since the test function now has both an explicit and implicit dependence on $q$, one must apply the chain rule in order to evaluate the derivative
\begin{align*}
&\int d^{4}p' \, d^{4}p \ \delta_{0}^{(+)}\!(p) \, \frac{\partial}{\partial p'_{\mu}} \delta^{4}(p'-p) \, f(p',p)  \\
&=-2\pi \!\int d^{3}\bar{\uvec{p}} \, \Bigg[  \frac{\partial}{\partial q_{\mu}}\!\left(\tfrac{1}{2}\sqrt{(\bar{\uvec{p}} -\tfrac{1}{2}\uvec{q})^{2}}    \right)^{\!\! -1} \bar{f}\!\left(\bar{p}^{0}_{\star}, \bar{\uvec{p}}, q \right)  + \frac{1}{2|\bar{\uvec{p}}|}\frac{d \bar{p}^{0}_{\star}}{dq_{\mu}} \frac{\partial \bar{f}\!\left(\bar{p}^{0}, \bar{\uvec{p}}, q \right)}{\partial \bar{p}^{0}}  + \frac{1}{2|\bar{\uvec{p}}|} \frac{\partial \bar{f}\!\left(\bar{p}^{0}_{\star}, \bar{\uvec{p}}, q \right)}{\partial q_{\mu}}  \Bigg]_{\!q=0} \\
&= 2\pi \!\int d^{3}\bar{\uvec{p}} \, \Bigg[ \frac{g^{\mu k}\bar{p}_{k}}{4|\bar{\uvec{p}}|^{3}}\bar{f}\!\left(\bar{p}, 0 \right) -\frac{\bar{p}^{\mu}}{(2|\bar{\uvec{p}}|)^{2}} \frac{\partial \bar{f}\!\left(\bar{p}, 0 \right)}{\partial \bar{p}^{0}}    - \frac{1}{2|\bar{\uvec{p}}|} \frac{\partial \bar{f}\!\left(\bar{p}, 0 \right)}{\partial q_{\mu}} \Bigg]_{\bar{p}^{0}= |\bar{\uvec{p}}|} \\
&= \int d^{4}\bar{p} \, d^{4}q  \, \Bigg[\frac{\bar{p}^{\mu}}{2\bar{p}^{0}}\frac{\partial}{\partial\bar{p}^{0}} \delta_{0}^{(+)}\!(\bar{p}) \, \delta^{4}(q) + \delta_{0}^{(+)}\!(\bar{p}) \, \frac{\partial}{\partial q_{\mu}}\delta^{4}(q)  \Bigg] \bar{f}(\bar{p},q),
\end{align*}
which proves the equality in Eq.~\eqref{dist_rel1}. \\

\noindent
In both of the GFF constraint calculations one is required to evaluate the product of delta-derivatives with specific components of the EMT matrix element. This amounts to understanding how these delta-derivatives act on the coefficients $F(\bar{p},q)$ of the various GFFs. Since these coefficients are continuous functions, one has the following identities:
\begin{align}
&\delta_{0}^{(+)}\!(p')\delta_{0}^{(+)}\!(p)\, F(\bar{p},q) \, \partial_{j}\delta^{3}(\uvec{q}) = \nonumber \\
& \hspace{10mm} (2\pi)\delta_{0}^{(+)}\!(\bar{p}) \left[ \frac{ F(\bar{p},0)}{2\bar{p}^{0}} \! \left( \partial_{j} - \frac{\bar{p}_{j}}{\bar{p}^{0}} \partial_{0}\right) \!\delta^{4}(q) -\frac{1}{2\bar{p}^{0}} \left(\frac{\partial F}{\partial q^{j}} -\frac{\bar{p}_{j}}{\bar{p}^{0}} \frac{\partial F}{\partial q^{0}}\right)_{\!\! q=0}  \!\delta^{4}(q)    \right], \label{D_dist2} \\
&\delta_{0}^{(+)}\!(p')\delta_{0}^{(+)}\!(p)\, F(\bar{p},q) \, \partial_{j}\partial^{k}\delta^{3}(\uvec{q}) =  \nonumber \\
& \hspace{5mm} (2\pi)\delta_{0}^{(+)}\!(\bar{p}) F(\bar{p},0)\left[ -\frac{1}{2(\bar{p}^{0})^{2}}\left( \bar{p}^{k}\frac{\partial}{ \partial q^{j}} +\bar{p}_{j}\frac{\partial}{\partial q_{k}} \right)\frac{\partial}{\partial q^{0}} +\frac{\bar{p}^{k}\bar{p}_{j}}{2(\bar{p}^{0})^{3}}\frac{\partial}{\partial q^{0}}\frac{\partial}{\partial q_{0}}     
+\frac{1}{2\bar{p}^{0}}\frac{\partial}{\partial q_{k}}\frac{\partial}{\partial q^{j}}  \right]_{\! q=0} \!\delta^{4}(q)  \nonumber \\
&\hspace{5mm} +(2\pi)\frac{F(\bar{p},0)}{8(\bar{p}^{0})^{2}}\left[g^{k}_{\ j} + \frac{\bar{p}^{k}\bar{p}_{j}}{(\bar{p}^{0})^{2}}  \right] \frac{\partial}{\partial \bar{p}^{0}} \delta_{0}^{(+)}\!(\bar{p}) \delta^{4}(q)   \nonumber \\
& \hspace{5mm} +(2\pi)\delta_{0}^{(+)}\!(\bar{p}) \left[ -\frac{\bar{p}^{k}}{2(\bar{p}^{0})^{2}}\frac{\partial^{2} F}{\partial q^{0} \partial q^{j}} -\frac{\bar{p}_{j}}{2(\bar{p}^{0})^{2}}\frac{\partial^{2} F}{\partial q^{0} \partial q_{k}} +\frac{\bar{p}^{k}\bar{p}_{j}}{2(\bar{p}^{0})^{3}}\frac{\partial^{2} F}{\partial q^{0} \partial q^{0}}   
+\frac{1}{2\bar{p}^{0}}\frac{\partial^{2} F}{\partial q_{k} \partial q^{j}}  \right]_{\! q=0}  \!\delta^{4}(q) \nonumber \\
&\hspace{5mm} +(2\pi)\delta_{0}^{(+)}\!(\bar{p}) \left[ \frac{\bar{p}^{k}}{2(\bar{p}^{0})^{2}}\frac{\partial F}{\partial q^{0}} - \frac{1}{2\bar{p}^{0}}\frac{\partial F}{\partial q_{k}}   \right]_{\! q=0} \!\partial_{j}\delta^{4}(q) +(2\pi)\delta_{0}^{(+)}\!(\bar{p}) \left[ \frac{\bar{p}_{j}}{2(\bar{p}^{0})^{2}}\frac{\partial F}{\partial q^{0}} - \frac{1}{2\bar{p}^{0}}\frac{\partial F}{\partial q^{j}}   \right]_{\! q=0} \!\partial^{k}\delta^{4}(q) \nonumber \\
&\hspace{5mm} +(2\pi)\delta_{0}^{(+)}\!(\bar{p}) \left[ \frac{\bar{p}^{k}}{2(\bar{p}^{0})^{2}}\frac{\partial F}{\partial q^{j}} + \frac{\bar{p}_{j}}{2(\bar{p}^{0})^{2}}\frac{\partial F}{\partial q_{k}} - \frac{\bar{p}^{k}\bar{p}_{j}}{(\bar{p}^{0})^{3}}\frac{\partial F}{\partial q^{0}}  \right]_{\! q=0} \!\partial^{0}\delta^{4}(q).
\label{dd_dist}
\end{align}
Both of these relations are proven in a similar manner to Eq.~\eqref{dist_rel1}, except in Eq.~\eqref{dd_dist} one has the added complication of having two nested derivatives, which introduces a significant number of additional contributions.

\bibliographystyle{JHEP}

\bibliography{refs}

\end{document}